\documentclass[12pt]{article}
\usepackage{graphicx}
\usepackage{hyperref}

\def \b{{\cal B}}
\def \bea{\begin{eqnarray}}
\def \beq{\begin{equation}}
\def \eea{\end{eqnarray}}
\def \eeq{\end{equation}}
\def \ob{\overline{B}^0}
\def \obs{\overline{B}^0_s}
\def \od{\overline{D}^0}

\begin{document}
\rightline{EFI 12-10}
\rightline{arXiv:1205.4964v3}
\rightline{July 2012}

\bigskip
\centerline{\bf PROSPECTS FOR IMPROVED $\Lambda_c$ BRANCHING FRACTIONS}
\bigskip

\centerline{Jonathan L. Rosner}
\centerline{\it Enrico Fermi Institute and Department of Physics}
\centerline{\it University of Chicago, 5620 S. Ellis Avenue, Chicago, IL 60637}
\bigskip

\begin{quote}
The experimental uncertainty on the branching fraction $\b(\Lambda_c \to p K^-
\pi^+) = (5.0 \pm 1.3)\%$ has not decreased since 1998, despite a much larger
data sample.  Uncertainty in this quantity dominates that in many other
quantities, including branching fractions of $\Lambda_c$ to other modes,
branching fractions of $b$-flavored baryons, and fragmentation fractions of
charmed and bottom quarks.  Here we advocate a lattice QCD
calculation of the form factors in $\Lambda_c \to \Lambda \ell^+ \nu_\ell$
(the case $\ell = e^+$ is simpler as the mass of the lepton can be neglected).
Such a calculation would yield an absolute prediction for the rate for
$\Lambda_c \to \Lambda \ell^+ \nu_\ell$.  When combined with the $\Lambda_c$
lifetime, it could provide a calibration for an improved set of $\Lambda_c$
branching fractions as long as the accuracy exceeds about 25\%.
\end{quote}

\leftline{PACS numbers: 14.20.Lq, 14.65.Dw, 13.30.Ce, 12.38.Gc}
\bigskip

\section{INTRODUCTION}

Despite the accumulation of a vastly greater sample of charmed particles
in $e^+ e^-$, $ep$, and hadron-hadron collisions, the most accurately known
branching fraction for the decay of the lowest-lying charmed baryon
$\Lambda_c$, $\b(\Lambda_c \to p K^- \pi^+) = (5.0 \pm 1.3)\%$, has
remained at the same value since 1998.  It was only pinned down to that
accuracy thanks to constructive suggestions by Dunietz \cite{Dunietz:1998uz}.
This branching fraction sets the scale for a number of other quantities
which depend on it.  Many other $\Lambda_c$ branching fractions are measured
through their ratio to the $p K^- \pi^+$ mode \cite{Nakamura:2010}.  It sets
the scale for $b$-flavored baryon branching fractions, and governs
fragmentation fractions of charm and bottom quarks into baryons.

In the present paper we advocate improvement of accuracy of the semileptonic
branching fraction $\b(\Lambda_c \to \Lambda e^+ \nu_e)$, whose current
value is $(2.1 \pm 0.6)\%$, via a lattice QCD calculation of the relevant
form factors.  Such calculations have been performed for the semileptonic
decays of charmed mesons, $D \to K \ell \nu_\ell$ and $D \to \pi \ell \nu_\ell$
\cite{lattice}, which are characterized by two form factors.  Although four
form factors are relevant to $\Lambda_c \to \Lambda \ell \nu_\ell$ in the limit
of zero lepton mass, the difficulty of such a calculation is outweighed by its
importance.  A calculation enabling the prediction of the rate for $\Lambda_c
\to \Lambda e^+ \nu_e$ (and hence its branching fraction, given
$\tau(\Lambda_c) = 200 \pm 6$ fs \cite{Nakamura:2010}) to an accuracy of better
than about 25\% would represent an improvement on a wide variety of key
quantities.

In Section II we review various quantities which could profit from improvement
in the accuracy of $\b(\Lambda_c \to \Lambda e^+ \nu_e)$.  We discuss in
Section III the present status of understanding of form factors in this
decay.  The corresponding semileptonic decay $\Lambda_b \to \Lambda_c e^-
\bar \nu_e$, to which the Heavy Quark Effective Theory (HQET) can be applied,
is treated in Section IV.  Some remarks are made in Section V regarding the
``calibrating'' mode $\Lambda_c \to p K^- \pi^+$, while Section V
concludes.

\section{DEPENDENT QUANTITIES}

\subsection{$\Lambda_c$ branching fractions}

Many $\Lambda_c$ branching fractions are determined by their ratio with
respect to $\b(\Lambda_c \to p K^- \pi^+)$ \cite{Nakamura:2010}:  For example,
\beq \label{eqn:lcrat}
\frac{\b(\Lambda_c \to \Lambda e^+ \nu_e)}{\b(\Lambda_c \to p K^- \pi^+)} =
0.41 \pm 0.07~;~~ \frac{\b(\Lambda_c \to \Lambda \pi^+)}{\b(\Lambda_c \to
p K^- \pi^+)} = 0.204 \pm 0.019~.
\eeq
In the last quantity we use the Particle Data group ``average.''
As $\b(\Lambda_c \to
p K^- \pi^+) = (5.0 \pm 1.3)\%$ is known to only a fractional error of 26\%,
this limits the accuracy to which quantities depending on it can be determined.
Other ratios \cite{Nakamura:2010} are
\beq \label{eqn:lcrat1}
\frac{\b(\Lambda_c \to \Lambda \pi^+ \pi^+ \pi^-)}{\b(\Lambda_c \to p K^-
\pi^+)} = 0.522 \pm 0.032~;~~ \frac{\b(\Lambda_c \to p \bar K^0)}
{\b(\Lambda_c \to p K^- \pi^+)} = 0.47 \pm 0.04~,
\eeq
using ``average'' values in both cases.  We advocate instead making a modest
improvement in the first ratio of Eq.\ (1) and
calibrating $\Lambda_c$ branching fractions by the $\Lambda e^+ \nu_e$ mode.

\subsection{$\Lambda_b$ branching fractions}

Most tabulated $\Lambda_b$ branching fractions involve a $\Lambda_c$ in the
final state \cite{Nakamura:2010}.  (An exception is the recently observed
decay $\Lambda_b \to \Lambda \mu^+ \mu^-$ \cite{Aaltonen:2011qs}.)  Examples
are
$$
\b(\Lambda_b \to \Lambda_c^+ \ell^- \bar \nu_\ell) = 0.050^{+0.011+0.016}
_{-0.008-0.012}
$$
\cite{Abdallah:2003gn},
\beq
\b(\Lambda_b \to \Lambda_c^+ \pi^-)
= (8.8\pm2.8\pm1.5)\times 10^{-3}
\eeq
\cite{Abulencia:2006df}.
The former measurement makes use of the branching fractions of $\Lambda_c$
to $p K^- \pi^+$, $\Lambda \pi^+ \pi^+ \pi^-$, and $p K_S$, while the latter
employs only the $p K^- \pi^+$ mode.  As $\b(\Lambda \pi^+ \pi^+ \pi^-)$ and
$\b(p K_S)$ are quoted with respect to $\b(p K^- \pi^+)$, their accuracies are
limited as well.

\subsection{Fragmentation fractions}

Individual probabilities for $c \to (D^0, D^+, D_s^+, \Lambda_c, \ldots)$ do
not seem to have been quoted in the literature.  However, the corresponding
fractions for $ b \to (\ob, B^-, \obs, \Lambda_b)$ are noted in the
Particle Data Group's mini-review on $B^0$--$\bar B^0$ mixing 
\footnote{In Ref.\ \cite{Nakamura:2010} see the mini-review by Schneider,
pp.\ 973--980, Table 1.}
and in a recent study by the LHCb Collaboration \cite{Aaij:2011jp}.  Such
fractions are needed in a wide variety of applications, including the
interpretation of CP asymmetries in same-sign dimuon production
at the Tevatron \cite{Abazov:2010hv}, and in studies of top quark production.

As an illustration of the uncertainty associated with $\Lambda_c$ branching
fractions, Ref.\ \cite{Aaij:2011jp} finds the ratio of strange $B$ meson to
light $B$ meson ($\ob,B^-$) production to be
\beq
\frac{f_s}{f_u + f_d} = 0.134 \pm 0.004^{+0.011}_{-0.010}~,
\eeq
but a much larger error in the ratio of $\Lambda_b$ to light meson production:
$$
\frac{f_{\Lambda_b}}{f_u + f_d}=[0.404 \pm 0.017({\rm stat}) \pm 0.027({\rm
syst}) \pm 0.105({\rm Br})]
$$
\beq
\times[1 - (0.031 \pm 0.004({\rm stat}) \pm 0.003({\rm syst}))p_T({\rm GeV})]~.
\eeq
The uncertainty labeled ``Br'' is due to the 26\% uncertainty in the branching
fraction of $\Lambda_c$ to $p K^- \pi^+$.  An additional theoretical
uncertainty is associated with the assumption that the total semileptonic
widths of $\Lambda_b$ and light $B$ are equal up to a small correction $\xi$.
Denoting a generic charmed hadron by $D$, Ref.\ \cite{Aaij:2011jp} finds
\beq
\frac{f_{\Lambda_b}}{f_u + f_d}=\frac{n_{\rm corr}(\Lambda_b \to D \mu)}
{n_{\rm corr}(B \to \od \mu) + n_{\rm corr}(B \to D^- \mu)}
\frac{\tau_{B^-} + \tau_{B^-}}{2 \tau_{\Lambda_b}} (1 - \xi)~,
\eeq
quoting $\xi = (4 \pm 2)\%$.  Examples of results for $\xi$ using heavy-quark
and operator product expansions are $(2.1 \pm 0.6)\%$ \cite{Manohar:1993qn},
$(5.2 \pm 0.6)\%$ \cite{Jin:1997in}, and $\simeq 3\%$ \cite{Bigi:2011gf}.  A
simple kinematic model, by contrast, gives $\xi \simeq 11\%$
\cite{Gronau:2010if}.  This, parenthetically, emphasizes the importance of
measurement of the inclusive branching fraction $\b(\Lambda_b \to \ell^-
\bar \nu_\ell X)$, for which a value has never been quoted.  The inclusive
branching fraction $\b(\Lambda_c \to \ell^+ \nu_\ell X)$ is not particularly
well known either \cite{Nakamura:2010,Gronau:2010if}:
\beq
\frac{\Gamma(\Lambda_c \to e^+ \nu_e X)}{\bar \Gamma(D \to e^+ \nu_e X)} = 1.44
\pm 0.54
\eeq
[$\bar \Gamma$ denotes a $(D^0,D^+)$ average].  This ratio is
to be compared with the prediction of 1.67 in the model of Ref.\
\cite{Gronau:2010if} and about 1.2 based on a heavy-quark expansion including
$1/m_c^2$ terms \cite{Manohar:1993qn}.  It would be highly worthwhile to
improve the precision of these measurements, an effort well within the
capabilities of the BaBar and Belle Collaborations.

\section{FORM FACTORS IN $\Lambda_c \to \Lambda e^+ \nu_e$}

For a semileptonic
decay of one spin-1/2 hadron to another there are three vector and three
axial-vector form factors.  One of each is negligible in the limit of zero
lepton mass, which we shall assume.  There remain two vector and two
axial-vector form factors, but for an arbitrary semileptonic decay $\Lambda_1
\to \Lambda_2 \ell \nu_\ell$ in the heavy-quark limit of $\Lambda_1$
all form factors appear multiplying a factor $1 - \gamma_5$ and hence
the vector and axial-vector form factors are equal pairwise.  The weak current
matrix element then may be written \cite{Korner:1991ph} as
\beq
\langle \Lambda_2 | J_\mu^{V+A} | \Lambda_1 \rangle = \bar u(P_2) [f_1(q^2)
\gamma_\mu (1-\gamma_5) + f_2(q^2) v_1 \!\!\!\!\!/~(1-\gamma_5)]u(P_1)~,
\eeq
We have denoted the (initial,final) $\Lambda$ by $\Lambda_{(1,2)}$ with
four-momentum $P_{(1,2)}$, mass $M_{(1,2)}$, and covariant four-velocity
$v_{(1,2)} = P_{(1,2)}/M_{(1,2)}$.  The four-momentum transfer to the lepton
pair is $q = P_1 - P_2$.  (In the
heavy-quark limit for the final $\Lambda$, $f_2 = 0$ and $f_1 = 1$ at $q^2 =
q_{\rm max}^2$.)  The form factors are assumed to be in a constant ratio
$r=f_2/f_1$, and to be governed by a dipole structure in $q^2$.  With the
choice of the $D_s^*$ mass in the dipole form factor, the rate for $\Lambda_c
\to \Lambda e^+ \nu_e$ is then predicted to be
\beq
\Gamma(\Lambda_c \to \Lambda e^+ \nu_e) = \left\{ \begin{array}{l}
1.57 \times 10^{11}~{\rm s}^{-1}~~{\rm for}~r = 0~, \cr
1.90 \times 10^{11}~{\rm s}^{-1}~~{\rm for}~r = -0.25~, \end{array} \right.
\eeq
where the latter value is preferred on the basis of an expansion in the
inverse of the strange quark mass (admittedly a crude approximation).

Experimental information on the decay $\Lambda_c \to \Lambda e^+ \nu_e$ comes
from the ARGUS \cite{Albrecht:1991bu} and CLEO \cite{Bergfeld:1994gt}
Collaborations:
\beq \label{eqn:ratio}
\frac{\Gamma(\Lambda_c \to \Lambda e^+ \nu_e)}
     {\Gamma(\Lambda_c \to p K^-\pi^+)}
= \left\{ \begin{array}{l} 0.38 \pm 0.14~\cite{Albrecht:1991bu} \cr
                           0.42 \pm 0.07~\cite{Bergfeld:1994gt} \cr
                           0.41 \pm 0.07~\cite{Nakamura:2010} \end{array}
\right.~.
\eeq
Combining the last of these (the PDG average) with $\b(\Lambda_c \to p K^-
\pi^+) = (5.0 \pm 1.3)\%$ \cite{Nakamura:2010} and the $\Lambda_c$ lifetime
$\tau(\Lambda_c) = (200 \pm 6)$ fs \cite{Nakamura:2010} one finds the
experimental value
\beq \label{eqn:lcr}
\Gamma(\Lambda_c \to \Lambda e^+ \nu_e) = (1.03 \pm 0.32)\times 10^{11}
~{\rm s}^{-1}~,
\eeq
somewhat below the predictions of Ref.\ \cite{Korner:1991ph}.

To give a qualitative idea of the expected shape of the leading form factor
(the one which does not vanish in the limit of heavy initial and final quarks
in the $\Lambda$), we adapt a discussion of the decay $\Lambda_b \to
\Lambda_c e^- \bar \nu_e$ \cite{Leibovich:2003tw} to the case of $\Lambda_c \to
\Lambda e^+ \nu_e$, treating the strange quark in the $\Lambda$ as ``heavy''.
(As shown, for example, in Ref.\ \cite{Amundson:1992xp} for charmed meson
semileptonic decays, this assumption has limited validity.)

The Isgur-Wise \cite{IW} variable $w = v_1 \cdot v_2$ is related to $q^2$ by
\beq
w = \frac{M_1^2 + M_2^2 - q^2}{2 M_1 M_2}
\eeq
and is equal to 1 at the zero-recoil point $q^2 = Q_{\rm max}^2=(M_1 - M_2)^2$.

\begin{figure}
\begin{center}
\includegraphics[width=0.6\textwidth]{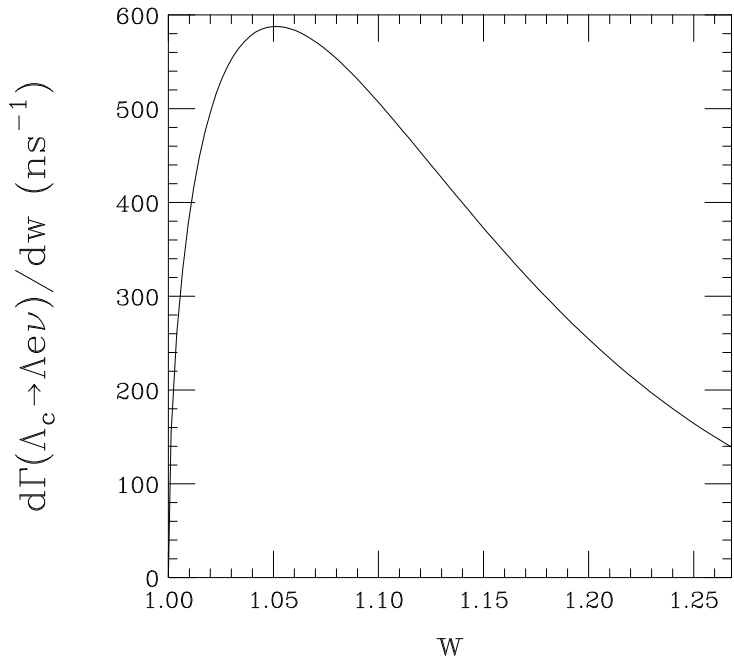}
\end{center}
\caption{Differential decay rate for $\Lambda_c \to \Lambda \ell \nu_\ell$
with respect to Isgur-Wise variable $w$, with the Isgur-Wise function
$\zeta(w) = \exp[-4.75(w-1)]$ reproducing the central value of the observed
decay rate (\ref{eqn:lcr}).
\label{fig:lcl}}
\end{figure}

The differential decay rate in the heavy-quark limit and the limit of vanishing
final lepton mass is \cite{Leibovich:2003tw}
\beq
\frac{d \Gamma(\Lambda_1 \to \Lambda_2 \ell \nu_\ell)}{d w} = \frac{G_F^2 M_1^5 
|V_{ij}|^2}{12 \pi^3} r^3 \sqrt{w^2 - 1}
[3w(1+r^2) - 2r(1+2w^2)]~\zeta(w)^2~,
\eeq
where $V_{ij}$ is the appropriate CKM matrix element for the semileptonic
quark transition $i \to j \ell \nu_\ell$, $r \equiv M_2/M_1$, and $\zeta(w)$
is the Isgur-Wise function, normalized to $\zeta(1) = 1$.  A simple form
which we shall adopt is $\zeta(w) = \exp[-\rho^2(w-1)]$.  Taking
\cite{Nakamura:2010} $M_1 = 2.28646$ GeV, $M_2 = 1.115683$ GeV, $|V_{cs}| =
0.97343$, we find the central value of Eq.\ (\ref{eqn:lcr}) is reproduced
for $\rho^2 = 4.75$.  The corresponding spectrum for $d\Gamma/dw$ is shown
in Fig.\ \ref{fig:lcl}.  A similar shape is to be expected
for a realistic lattice gauge theory calculation, which should take into
account the contributions of form factors which vanish in the heavy-quark
limit.

The lattice calculation of form factors in $\Lambda_c \to\Lambda \ell \nu_\ell$
may prove to be quite challenging.  For $D \to K \ell
\nu_l$, errors in form factors of several percent have been achieved
\cite{lattice}.  One could hope for the baryonic case to be similar with
the replacement of a light antiquark spectator in $D \to K$ by a $ud$
diquark with $I=J=0$ in $\Lambda_c \to \Lambda$.  However, the $ud$ diquark
can undergo internal excitations, making the situation more complicated than
in the mesonic case.  A note of caution is also provided by the current
status of the lattice calculation of semileptonic $\Lambda_b$ decays, which
we now discuss briefly.

\section{FORM FACTORS IN $\Lambda_b \to \Lambda_c e^- \bar \nu_e$}

\begin{figure}
\begin{center}
\includegraphics[width=0.6\textwidth]{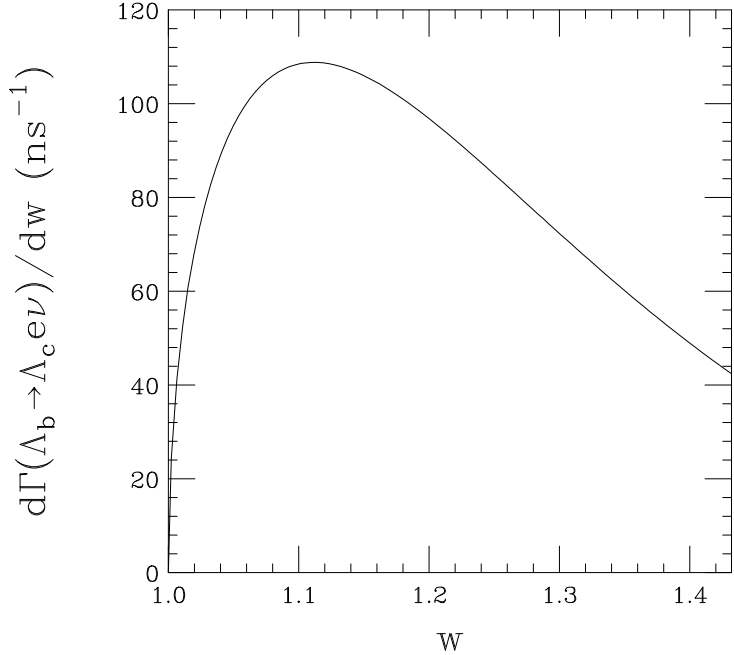}
\end{center}
\caption{Differential decay rate for $\Lambda_b \to \Lambda_c \ell \nu_\ell$
with respect to Isgur-Wise variable $w$, with the Isgur-Wise function
$\zeta(w) = \exp[-2.3(w-1)]$ reproducing the central value of the observed
decay rate (\ref{eqn:lblc}).
\label{fig:lblc}}
\end{figure}

The calculation of the previous section can be adapted to the decay $\Lambda_b
\to \Lambda_c \ell \nu_\ell$, for which the heavy-quark limit should be a
better approximation.  We take \cite{Nakamura:2010} $M_1 = 5.6202$ GeV, $M_2 =
2.28646$ GeV, and $|V_{cb}| = 0.041$.  The experimental branching fraction
is $\b(\Lambda_b \to \Lambda_c e^- \bar \nu_e) = (5.0^{+1.9}_{-1.4})\%$;
combined with the $\Lambda_b$ lifetime $\tau(\Lambda_b) = (1.425 \pm 0.032)
\times 10^{-12}$ s, this gives a decay rate
\beq \label{eqn:lblc}
\Gamma(\Lambda_b \to \Lambda_c e^- \bar \nu_e) = (3.5^{+1.3}_{-1.0}) \times
10^{10}~{\rm s}^{-1}~,
\eeq
whose central value is reproduced with the choice $\rho^2 = 2.3$ in the
Isgur-Wise function.  A similar though not identical result is obtained
by the DELPHI Collaboration \cite{Abdallah:2003gn}.  The corresponding
differential decay rate is shown in Fig.\ \ref{fig:lblc}.

There exists a lattice QCD study of the decay $\Lambda_b \to \Lambda_c e^- \bar
\nu_e$ \cite{Gottlieb:2003yb}.  The function $\zeta(w)$, if normalized to 1
at $w=1$, is seen to fall to $0.65 \pm 0.03$ at $w = 1.06$, corresponding
to $\rho^2 = 7.2 \pm 0.8$.  This is quite far from the value which reproduces
the observed decay rate.
It is not clear whether this is an intrinsic shortcoming of the lattice
approach, which would bode poorly for calculating
$\Gamma(\Lambda_c \to \Lambda e^+ \nu_e)$ to better than 25\%, or a feature
of the specific calculation which might be improved using more recent
techniques.

\section{REMARKS ON THE MODE $\Lambda_c \to p K^- \pi^+$}

The decay $\Lambda_c \to \Lambda e^+ \nu_e$ has one disadvantage with respect
to all-hadronic modes such as $\Lambda_c \to p K^- \pi^+$:  In the semileptonic
decay, one must ensure that nothing besides the neutrino is missing, whereas
an all-charged mode such as $p K^- \pi^+$ provides a useful kinematic
constraint.  It is therefore worth reviewing briefly the ingredients in the
present determination of the ``calibrating'' branching fraction $\b(\Lambda_c
\to p K^- \pi^+) = (5.0 \pm 1.3)\%$ \footnote{See Burchat, mini-review in
Ref.\ \cite{Nakamura:2010}, pp.\ 1260--1261.}, to see if some improvement
in that quantity is possible.

One determination of $\b(\Lambda_c \to p K^- \pi^+) = (5.0 \pm 1.3)\%$ is
obtained by averaging two types of measurements.  The first measures a
combined branching ratio $\b(\bar B \to \Lambda_c X) \cdot \b(\Lambda_c \to
p K^- \pi^+)$ and estimates the first factor by assuming that $\Lambda_c X$
final states other than $\Lambda_c \overline{N} X$ are negligible.  This
assumption was called into question in Ref.\ \cite{Dunietz:1998uz}.  The
second relies upon measurement of the ratio (\ref{eqn:ratio}) and the
assumptions that (i) the semileptonic decay of $\Lambda_c$ is saturated
by the $\Lambda e^+ \nu_e$ final state, and (ii) all inclusive semileptonic
decay rates of charmed particles are equal.  While this appears to be true
for mesons, it is far from established in the case of $\Lambda_c$
\cite{Gronau:2010if}.

An independent determination of $\b(\Lambda_c \to p K^- \pi^+) = (5.0 \pm
0.5 \pm 1.2)\%$ was performed by the CLEO Collaboration \cite{Jaffe:2000nw}.
It analyzes $e^+ e^- \to c \bar c \to \bar D \bar p X$ continuum events, where
the $\bar c$ is tagged by the presence of the $\bar D$, the $\bar p$ is in the
hemisphere opposite to the $\bar D$ (to reduce non-signal background), and it
is assumed that there is always a $\Lambda_c$ present in $X$ to compensate for
charm and baryon number.  One then measures the $p K^- \pi^+$ yield in the same
hemisphere as the $\bar p$ to obtain $\b(\Lambda_c \to p K^- \pi^+)$.
Backgrounds against which one has to guard include $D \bar D N \bar p$ and
kaons producing fake antiproton tags.

The measurement of Ref.\ \cite{Jaffe:2000nw} is based on 3.1 fb$^{-1}$
collected at the $\Upsilon(4S)$ resonance and 1.6 fb$^{-1}$ collected about
60 MeV below it, corresponding to about 5 million continuum $c \bar c$ events.
Although the experimental error is dominated by systematics, the authors
note that more data would allow better understanding of backgrounds such as
$D \bar D N \bar p$.  It would be worth seeing how well one could perform
such an analysis with the much larger data samples available to the BaBar
and Belle Collaborations.

\section{CONCLUSIONS}

The importance of improved knowledge of the decay rate for $\Lambda_c \to
\Lambda e^+ \nu_e$ has been stressed.  Progress is possible in principle upon
a variety of fronts, including (1) lattice gauge theory calculations of form
factors, (2) improved measurements of ratios of $\Lambda_c$ branching
fractions, (3) improved determination of inclusive $\Lambda_c$ and $\Lambda_b$
semileptonic branching fractions, and (4) validation of lattice QCD
calculations and heavy-quark symmetry through the continued study of $\Lambda_b
\to \Lambda_c \ell \nu_\ell$.  Many quantities depend upon an absolute
calibration of $\Lambda_c$ branching fractions, a goal whose attainment is
long overdue.

\section*{ACKNOWLEDGMENTS}

This work was initiated at a Heavy Flavors Workshop at the University of
Washington.  I am grateful to Ann Nelson for the invitation to attend, and to
S. Gottlieb, Z. Ligeti, S. Stone, and R. Van de Water for helpful discussions.
This work was supported in part by the United States Department of Energy under
Grant No.\ DE-FG02-90ER40560.


\begin{thebibliography}{99}

\bibitem{Dunietz:1998uz} 
  I.~Dunietz,
  Phys.\ Rev.\ D {\bf 58}, 094010 (1998).

\bibitem{Nakamura:2010}
K. Nakamura {\it et al.} (Particle Data Group), J. Phys. G {\bf 37}, 075021
(2010), and partial 2011 update for the 2012 edition.

\bibitem{lattice} See, for example, J. A. Bailey {\it et al.} (Fermilab
Lattice and MILC Collaborations),
Proc.\ Sci., LATTICE2011 (2011) 270; 
C. Davies, {\it ibid.},
Proc.\ Sci., LATTICE2011 (2011) 019.

\bibitem{Aaltonen:2011qs} 
  T.~Aaltonen {\it et al.} (CDF Collaboration),
  Phys.\ Rev.\ Lett.\  {\bf 107}, 201802 (2011).

\bibitem{Abdallah:2003gn}
  J.~Abdallah {\it et al.} (DELPHI Collaboration),
  Phys.\ Lett.\ B {\bf 585}, 63 (2004).

\bibitem{Abulencia:2006df}
  A.~Abulencia {\it et al.} (CDF Collaboration),
  Phys.\ Rev.\ Lett.\ {\bf 98}, 122002 (2007).

\bibitem{Aaij:2011jp}
  R.~Aaij {\it et al.} (LHCb Collaboration),
  Phys.\ Rev.\ D {\bf 85}, 032008 (2012).

\bibitem{Abazov:2010hv} 
  V.~M.~Abazov {\it et al.} (D0 Collaboration),
  Phys.\ Rev.\ D {\bf 82}, 032001 (2010);
  Phys.\ Rev.\ Lett.\  {\bf 105}, 081801 (2010);
  [arXiv:1007.0395 [hep-ex]];
  Phys.\ Rev.\ D {\bf 84}, 052007 (2011).

\bibitem{Manohar:1993qn} 
  A.~V.~Manohar and M.~B.~Wise,
  Phys.\ Rev.\ D {\bf 49}, 1310 (1994).

\bibitem{Jin:1997in}
  C.~Jin,
  Phys.\ Rev.\ D {\bf 56}, 7267 (1997).

\bibitem{Bigi:2011gf} 
  I.~I.~Bigi, T.~Mannel and N.~Uraltsev,
  J. High Energy Phys.\ 09 (2011) 012.

\bibitem{Gronau:2010if} 
  M.~Gronau and J.~L.~Rosner,
  Phys.\ Rev.\ D {\bf 83}, 034025 (2011).

\bibitem{Korner:1991ph} 
  J.~G.~Korner and M.~Kramer,
  Phys.\ Lett.\ B {\bf 275}, 495 (1992).

\bibitem{Albrecht:1991bu} 
  H.~Albrecht {\it et al.} (ARGUS Collaboration),
  Phys.\ Lett.\ B {\bf 269}, 234 (1991).

\bibitem{Bergfeld:1994gt} 
  T.~Bergfeld {\it et al.} (CLEO Collaboration),
  Phys.\ Lett.\ B {\bf 323}, 219 (1994).

\bibitem{Leibovich:2003tw} 
  A.~K.~Leibovich, Z.~Ligeti, I.~W.~Stewart and M.~B.~Wise,
  Phys.\ Lett.\ B {\bf 586}, 337 (2004).

\bibitem{Amundson:1992xp}
  J.~F.~Amundson and J.~L.~Rosner,
  Phys.\ Rev.\ D {\bf 47}, 1951 (1993).

\bibitem{IW}
  N. Isgur and M. B. Wise, Phys.\ Lett.\ B {\bf 232}, 113 (1989);
  {\bf 237}, 527 (1990).

\bibitem{Gottlieb:2003yb} 
  S.~A.~Gottlieb and S.~Tamhankar,
  Nucl.\ Phys.\ Proc.\ Suppl.\  {\bf 119}, 644 (2003).

\bibitem{Jaffe:2000nw} 
  D.~E.~Jaffe {\it et al.} (CLEO Collaboration),
  Phys.\ Rev.\ D {\bf 62}, 072005 (2000).

\end{thebibliography}
\end{document}